\documentclass[acmtog,authorversion,balance=true]{acmart}

\copyrightyear{2023}
\acmYear{2023}
\setcopyright{rightsretained}
\acmConference[SA Conference Papers '23]{SIGGRAPH Asia 2023 Conference Papers}{December 12--15, 2023}{Sydney, NSW, Australia}
\acmBooktitle{SIGGRAPH Asia 2023 Conference Papers (SA Conference Papers '23), December 12--15, 2023, Sydney, NSW, Australia}
\acmDOI{10.1145/3610548.3618146}
\acmISBN{979-8-4007-0315-7/23/12}
\usepackage{amsmath} 
\usepackage{amsfonts}
\usepackage{booktabs} 
\usepackage{hyperref}
\usepackage[capitalize,nameinlink]{cleveref}

\usepackage{mathtools}
\usepackage[shortlabels]{enumitem}
\usepackage{soul}
\usepackage{xspace}
\usepackage{multirow}

\usepackage{graphicx} 
\usepackage{wrapfig}
\usepackage{pifont}

\usepackage{nicefrac}
\usepackage{tabularx} 

\usepackage[nolist]{acronym}
\usepackage{rotating} 

\ifdefined\TAPSversion
\else
    \usepackage{mathrsfs}
    \usepackage{tikz}
    \usepackage{overpic}
    \usepackage{pict2e}
    \usepackage{dsfont}
    \usepackage[most]{tcolorbox}
    \usepackage[outline]{contour}
\fi

\newcommand{\ie}{i.e.,\ }
\newcommand{\eg}{e.g.,\ }

\newcommand{\mymath}[2]{\newcommand{#1}{\TextOrMath{$#2$\xspace}{#2}}}
\mymath{\convolution}{*}
\mymath{\Kernel}{g}
\mymath{\KernelSpace}{\Kernel_\text{s}}
\mymath{\KernelTime}{\Kernel_\text{t}}
\mymath{\KernelTAA}{\Kernel_\text{a}}
\mymath{\KernelTimeCombined}{\Kernel_\text{ta}}
\mymath{\KernelTimeSustained}{\Kernel_{\text{t}_\text{sustained}}}
\mymath{\KernelTimeTransient}{\Kernel_{\text{t}_\text{transient}}}
\mymath{\KernelXSpace}{\Kernel_\text{x}}

\mymath{\FrameCount}{N}
\mymath{\Reference}{I}
\mymath{\ReferenceSequence}{\mathbf{\Reference}}
\mymath{\Image}{Q}
\mymath{\ImageSequence}{\mathbf{\Image}}
\mymath{\Estimate}{R}
\mymath{\EstimateSequence}{\mathbf{\Estimate}}
\mymath{\Sample}{u}
\mymath{\Samples}{S}
\mymath{\SamplesSequence}{\mathbf{\Samples}}
\mymath{\ErrorImage}{\epsilon}
\mymath{\ErrorImageSequence}{\mathbf{\ErrorImage}}
\mymath{\Error}{E}
\mymath{\ErrorSequence}{\mathbf{\Error}}
\mymath{\MCerror}{q}

\mymath{\ErrorImageSpectrum}{\hat{\ErrorImage}}
\mymath{\KernelSpectrum}{\hat{\Kernel}}

\mymath\temporalDomain{T}
\mymath\spatioTemporalDomain{XT}
\mymath\spatialDomain{X}
\mymath\spatialDomainX{X}
\mymath\spatialDomainY{Y}

\mymath{\FrequencySpatial}{\omega_\spatialDomain}
\mymath{\FrequencyTemporal}{\omega_\temporalDomain}
\mymath{\FrequencySpatialX}{\omega_X}
\mymath{\FrequencySpatialY}{\omega_Y}
\mymath{\FrequencyTemporalT}{\omega_T}

\mymath{\Point}{x}
\mymath{\PointCount}{n}

\mymath\Slice z
\mymath\SliceTemporal r
\mymath{\Density}{\mu}

\DeclareMathOperator*{\argmin}{arg\,min}

\newcommand{\dif}{\mathop{}\!\mathrm{d}}

\ifdefined\TAPSversion
\else
    \usepackage{xcolor}
    \usepackage{caption} 
    \usepackage{newfloat}  
    \usepackage{savesym}
    \usepackage{algorithm}
    \usepackage[noend]{algpseudocode}
    \savesymbol{Comment}
    \algrenewcommand\algorithmicindent{3mm}
    \algrenewcommand{\alglinenumber}[1]{\color{black!50}\fontsize{7.5}{6}\selectfont#1\color{black!30}:\phantom{*}}

    \tcbset{
        colback=white!98!black,
        colframe=white!75!black,
        boxrule=0.15mm,
        arc=0.3mm,
        boxsep=0pt,left=2pt,right=4pt,top=2pt,bottom=4pt
    }
\fi

\begin{acronym}
\acro{WoV}{Window of Visibility}
\acro{MC}{Monte Carlo}
\acro{EMA}{exponential moving average}
\acro{pRelMSE}{perceptual relative
mean squared error}
\end{acronym}

\newcommand{\pError}{\ac{pRelMSE}\xspace}

\ifdefined\TAPSversion
\else
    \DeclareFloatingEnvironment[
        name=Algorithm,
        placement=tbhp,
        within=none,
    ]{pseudocode}
    \crefname{pseudocode}{Alg.}{Algs.}
    \Crefname{pseudocode}{Algorithm}{Algorithms}
\fi

\definecolor{DarkGreen}{rgb}{0.0,0.6,0.0}
\DeclareFontFamily{U} {MnSymbolA}{}
\DeclareFontShape{U}{MnSymbolA}{m}{n}{
  <-6> MnSymbolA5
  <6-7> MnSymbolA6
  <7-8> MnSymbolA7
  <8-9> MnSymbolA8
  <9-10> MnSymbolA9
  <10-12> MnSymbolA10
  <12-> MnSymbolA12}{}
\DeclareFontShape{U}{MnSymbolA}{b}{n}{
  <-6> MnSymbolA-Bold5
  <6-7> MnSymbolA-Bold6
  <7-8> MnSymbolA-Bold7
  <8-9> MnSymbolA-Bold8
  <9-10> MnSymbolA-Bold9
  <10-12> MnSymbolA-Bold10
  <12-> MnSymbolA-Bold12}{}
\DeclareSymbolFont{MnSyA} {U} {MnSymbolA}{m}{n}
\DeclareMathSymbol{\rcurvearrowsw}{\mathrel}{MnSyA}{198}
\DeclareMathSymbol{\rcurvearrownw}{\mathrel}{MnSyA}{189}
\DeclareMathSymbol{\codeleftarrow}{\mathrel}{MnSyA}{2}

\newcommand{\AlgCommentTemplate}[2]{\hfill{\fontsize{6.7}{6}\selectfont\textcolor{DarkGreen}{\text{#1\;#2}}}}

\newcommand{\AlgCommentLeft}[1]{\AlgCommentTemplate{${\codeleftarrow}$}{#1}}

\DeclareGraphicsExtensions{.png,.jpg,.pdf,.ai,.psd}
\setcitestyle{nosort,square}

\begin{document}


\title{Perceptual error optimization for Monte Carlo animation rendering}

\author{Mi\v{s}a Kora\'{c}\,}
\authornote{These authors have contributed equally to this work.}
\email{korac@cg.uni-saarland.de}
\orcid{0009-0008-4390-5412}
\affiliation{%
    \institution{Saarland University, DFKI}
    \country{Germany}
}

\author{Corentin Sala\"{u}n\,}
\authornotemark[1]
\email{csalaun@mpi-inf.mpg.de}
\orcid{0000-0002-5112-7488}
\affiliation{%
    \institution{Max Planck Institute for Informatics}
    \country{Germany}
}

\author{Iliyan Georgiev}
\email{igeorgiev@adobe.com}
\orcid{0000-0002-9655-2138}
\affiliation{%
    \institution{Adobe}
    \country{UK}
}

\author{Pascal Grittmann}
\email{grittmann@cg.uni-saarland.de}
\orcid{0000-0002-5325-3744}
\affiliation{%
    \institution{Saarland University}
    \country{Germany}
}

\author{Philipp Slusallek}
\email{philipp.slusallek@dfki.de}
\orcid{0000-0002-2189-2429}
\affiliation{%
    \institution{Saarland University, DFKI}
    \country{Germany}
}

\author{Karol Myszkowski}
\email{karol@mpi-inf.mpg.de}
\orcid{0000-0002-8505-4141}
\affiliation{%
    \institution{Max Planck Institute for Informatics}
    \country{Germany}
}

\author{Gurprit Singh}
\email{gsingh@mpi-inf.mpg.de}
\orcid{0000-0003-0970-5835}
\affiliation{%
    \institution{Max Planck Institute for Informatics}
    \country{Germany}
}

\renewcommand{\shortauthors}{M.\,Kora\'{c}, C.\,Sala\"{u}n, I.\,Georgiev, P.\,Grittmann, P.\,Slusallek, K.\,Myszkowski, G.\,Singh}


\begin{abstract}

Independently estimating pixel values in Monte Carlo rendering results in a perceptually sub-optimal white-noise distribution of error in image space. Recent works have shown that perceptual fidelity can be improved significantly by distributing pixel error as blue noise instead. Most such works have focused on static images, ignoring the temporal perceptual effects of animation display. We extend prior formulations to simultaneously consider the spatial and temporal domains, and perform an analysis to motivate a perceptually better spatio-temporal error distribution. We then propose a practical error optimization algorithm for spatio-temporal rendering and demonstrate its effectiveness in various configurations.

\vspace{-1mm}

\end{abstract}




\begin{teaserfigure}
    \centering
    \vspace{-3mm}
    \hspace*{-1.4mm}
    \includegraphics[width=1.01\textwidth]{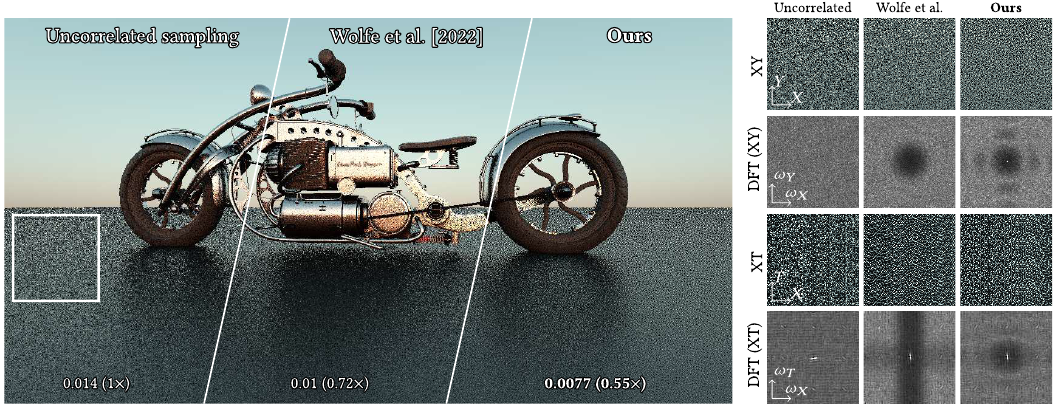}
    \vspace{-7mm}
    \caption{
        We propose an optimization framework to obtain perceptually pleasing error distribution in Monte Carlo animation rendering. The output of our algorithm is a sample set spanning multiple image pixels and frames. Here we show an image of a 30-frame sequence rendered with 1 sample/pixel per frame. We display a version of the animation filtered temporally using the kernel of~\citet{mantiuk2021fovvideovdp}, to mimic its perception at one time instant. On the right we a show spatial (XY) crop and a spatio-temporal (XT) slice, along with the power spectra (DFT) of their corresponding error images. Our error distribution exhibits better blue-noise properties than that of previous work~\cite{wolfe2022stbn}, also reflected in the perceptual error metric reported on the left (see \cref{sec:results}). To fully appreciate these results, please refer to the supplemental video and HTML viewer.
    }
    \label{fig:teaser}
    \vspace{2mm}
\end{teaserfigure}

\maketitle

\section{Introduction}

Monte Carlo rendering numerically estimates light-transport integrals via random sampling which causes visible noise in the resulting image. Much work has focused on combating this noise by reducing the error in each pixel individually, \eg via blue-noise or low-discrepancy sampling~\cite{singh2019analysis}. Applying such a pattern \emph{independently} within each pixel improves the convergence rate towards a noise-free result. However, the resulting white-noise distribution of error over the image is visually sub-optimal.

It is well understood in digital half-toning literature that the human visual system (HVS) is less sensitive to image error that has high-frequency, \ie blue-noise, distribution. \citet{georgiev2016blue} achieved such distribution in Monte Carlo rendering by carefully optimizing a global sample pattern \emph{across image pixels}. This pattern yields higher perceptual fidelity by making the pixel estimates as different from each other as possible. This improvement occurs because the HVS applies a low-pass filter to the image~\cite{chizhov2020perceptual}, and the negative pixel correlation effectively stratifies the input to the low-pass convolution.

Following the work of \citet{georgiev2016blue}, several practical methods have been devised to achieve high-quality blue-noise distribution for static-image rendering~\cite{heitz2019distributing,belcour2021bluenoise,ahmed2020screen}. These are mostly heuristically derived. An exception is the method of \citet{salaun2022scalable} which leverages the perceptual framework of \citet{chizhov2020perceptual} to compute a small sample set, tiled over the rendered image.

Reusing the same blue-noise sample set across the frames of an animation would maintain good blue-noise distribution, but the noise pattern would remain static over the image. This so-called shower-door effect \cite{kass2011coherent} degrades visual quality and disrupts the perception of motion. To address this problem, \citet{wolfe2022stbn} made a first attempt at obtaining an error distribution for animation rendering that is blue-noise in both image space and time. Lacking firm perceptual grounding, they extend existing blue-noise-mask algorithms to optimize separately across screen-space and time, which leads to visually suboptimal results.

In this paper, we combine the image-space model of \citet{chizhov2020perceptual} with a temporal perception model~\cite{mantiuk2021fovvideovdp} to quantify perceptual error in animation rendering and motivate the need for its high-frequency distribution in both space and time. We also incorporate explicit temporal filtering such as temporal anti-aliasing (TAA). Based on this spatio-temporal model, we adapt the optimization method of \citet{salaun2022scalable} to obtain scene-independent, precomputed sample sets. The resulting sample sets allow for low-sample animation rendering with higher perceptual fidelity than prior state of the art, thanks to the blue-noise distribution of error in both space and time. \Cref{fig:teaser} shows one frame of an animation rendered with our optimization algorithm.

\section{Related work}

Our goal is to optimize Monte Carlo rendering error \emph{across pixels} as blue noise, in both image space and time. The survey of \citet{singh2019analysis} discusses methods for achieving blue noise on one integration domain (\eg within a single pixel).

\paragraph{Blue-noise error distribution}
\label{sec:RelatedErrorDistribution}

Blue-noise distributions of image error appear frequently in dithering or stippling applications~\cite{ulichney1988dithering}. The reason for their use is the lower sensitivity of the HVS to high-frequency noise (``blue noise''), resulting in a less perceptible error. High-frequency noise distribution corresponds to negative correlation between pixel values in a neighbourhood. For Monte Carlo rendering, \citet{georgiev2016blue} proposed a first practical approach that optimizes a blue-noise sample mask via simulated annealing. Their approach is limited to low-dimensional integration with few samples. \citet{heitz2019low} addressed these limitations by optimizing the scrambling keys of a Sobol sequence~\cite{sobol1967}. \citet{belcour2021bluenoise} extended this optimization to a rank-1 lattice sampler. The method of \citet{ahmed2020screen} scrambles an image-space Sobol sequence according to a z-code ordering of pixels to achieve an approximate blue-noise distribution. \citet{salaun2022scalable} employed sliced optimal transport~\cite{paulin2020sliced} to obtain a sample set optimized according to the perceptual model of \citet{chizhov2020perceptual}. Recently, \citet{wolfe2022stbn} proposed extensions to the void-and-cluster~\cite{ulichney1988dithering} and \citeauthor{georgiev2016blue}'s~\shortcite{georgiev2016blue} algorithms to generate blue-noise sample masks for animation rendering. All the aforementioned methods are \emph{a priori}, \ie they compute scene-agnostic sample patterns. Such precomputation is beneficial for practical application, though superior quality can be achieved by tailoring the distribution to the specific image being rendered. This can be done through \emph{a posteriori} adaptation of sample distributions, once the pixels have been sampled~\cite{heitz2019distributing,chizhov2020perceptual}. We extend the image-space model of \citet{chizhov2020perceptual} to the temporal domain and apply the a priori optimization approach of \citet{salaun2022scalable} to acquire a sample pattern for each animation frame.

\paragraph{Perceptual modeling and rendering}

The contrast sensitivity function (CSF) is an important characteristic of the HVS that determines the threshold contrast that is perceivable in spatio-temporal signals. A vast majority of CSF measurements focus on spatial patterns \cite{Daly93,Barten:1999,Mantiuk2020}, and the resulting CSFs are modeled by a family of band-pass filters whose parameters change with luminance, color, and retinal eccentricity. Spatio-temporal CSFs have also been derived \cite{kelly79motion,DALY98,robson1966,mantiuk2022}, where temporal sensitivity \cite{delange1958} can be explained by sustained and transient temporal channels dedicated to processing slowly and quickly changing signals \cite{burbeck1980spatiotemporal,hammett1992two,mantiuk2021fovvideovdp}. The so-called window of visibility \cite{watson2013high,watson2016pyramid,Watson1986windowOfVisibility} is an example of such spatio-temporal CSF modeling. The window of visibility approach accounts for spatio-temporal signal processing and sampling that are inherent to any imaging pipeline. In rendering applications, such spatio-temporal CSFs have been used to focus expensive computation on the most visible regions only~\cite{myszkowski99perceptually,yee2001spatiotemporal}. In this work, we reduce perceived rendering error by employing a spatio-temporal CSF to optimize space-time sampling patterns.

\paragraph{Temporal anti-aliasing}

Temporal anti-aliasing (TAA) combines pixels across multiple frames to reduce noise~\cite{shinya1993spatial,schied2017spatiotemporal,schied2018gradient}. Such temporal filtering is simple and cheap, though ghosting artifacts arise if the scene changes too rapidly. These can be reduced via the use of motion vectors or other means of temporal reprojection~\cite{hanika2021fast}. Our method can optimize the perceived screen-space error distribution of a TAA-filtered animation.

\paragraph{Bounding integration error}

Quasi-Monte Carlo (QMC) integration methods use deterministic sample sequences. These sequences are carefully designed to minimize discrepancy which is a quality metric used to bound integration error~\cite{ermakov2019koksmaHlawka}. Recent work~\cite{paulin2020sliced} has shown an analogous error bound based on the Wasserstein distance instead \cite{kantorovich1958space, villani2008optimal}. This bound has been extended by \citet{salaun2022scalable} to perceptual error in single-image rendering. We further extend their bound to our spatio-temporal setting.

\begin{table}
    \centering
    \small
    \caption{
        Commonly used notations throughout the document.
    }
    \label{tab:Notation}
    \vspace{-1.5mm}
    \setlength{\tabcolsep}{2.5pt}
    \fontsize{6.9pt}{7pt}
    \begin{tabularx}{\columnwidth}{l X}
        \toprule
        \small \textbf{Notation} & \small\textbf{Description} \\
        \midrule
        $\Samples_i$, $\SamplesSequence = \{ \Samples_i \}$ & Sample set for frame $i$, sample set for entire frame sequence \\
        $\Estimate_i$, $\EstimateSequence = \{ \Estimate_i \}$ & Raw render result at frame $i$, sequence of all raw results \\
        $\Image_i$, $\ImageSequence =  \{ \Image_i \}$ & Displayed image at frame $i$, sequence of all displayed images \\
        $\Reference_i$, $\ReferenceSequence = \{ \Reference_i \}$ & Ground-truth image at frame $i$, sequence of all ground truths \\
        $\ErrorImage_i$, $\ErrorImageSequence = \{ \ErrorImage_i \}$ & Perceptual-error image at frame $i$, perceptual-error sequence \\
        $\KernelSpace$, $\KernelTime$ & Spatial perceptual kernel, temporal perceptual kernel \\
        $\KernelTAA$ & Explicit temporal accumulation (TAA) kernel \\
        \Density & Sample distribution (typically uniform) \\        
        \bottomrule
    \end{tabularx}
    \vspace{-1mm}
\end{table}

\section{Spatio-temporal perceptual model}
\label{secSpatioTemporalModel}

Our method builds on the perceptual model of \citet{chizhov2020perceptual}, which we extend to include the temporal model of \citet{mantiuk2021fovvideovdp} as well as explicit filtering via temporal anti-aliasing (TAA).

\paragraph{Notation}
Given a sequence $\ImageSequence = \{ \Image_i \}$ of rendered images, we aim to minimize their perceived error compared to the sequence of corresponding references $\ReferenceSequence = \{ \Reference_i \}$. Each image is a function $\Image_i(\Samples_i)$ of the sample pattern $\Samples_i$ that is used to render the $i$\textsuperscript{th} frame of an animation. We concisely express the sequence of rendered images $\ImageSequence(\SamplesSequence)$ as a function of the sequence of sample patterns. \Cref{tab:Notation} lists the most commonly used symbols throughout the paper.

\paragraph{Spatial perceptual error}
We follow \citet{chizhov2020perceptual} and model spatial perceptual response as a convolution. Hence the perceived error of the $i$\textsuperscript{th} frame viewed individually,
\begin{align}
    \label{eq:SpatialPerceptualError}
    \ErrorImage_i(\Samples_i)
        = \KernelSpace \convolution \Image_i(\Samples_i) - \KernelSpace \convolution \Reference_i 
        = \KernelSpace \convolution (\Image_i(\Samples_i) - \Reference_i)
        \text,
\end{align}
can be quantified by comparing the perceived image $\KernelSpace \convolution\Image_i$ to the perceived reference $\KernelSpace \convolution\Reference_i$. Here, $\KernelSpace$ is an image-space Gaussian kernel that approximates the human visual system's (HVS) point spread function~(PSF) \cite{chizhov2020perceptual}. The error image $\ErrorImage_i(\Samples_i)$ then measures the error for each pixel in the $i$\textsuperscript{th} frame. 

\paragraph{Spatio-temporal perceptual error}
\label{sec:SpatioTemporalError}

The human visual system (HVS) does not perceive each animation frame in isolation. Rather, it has been observed that temporal perception can be also modelled as a low-pass filter~\cite{mantiuk2021fovvideovdp,burbeck1980spatiotemporal,hammett1992two}. We incorporate temporal filtering with a kernel $\KernelTime$ into the spatial model~\eqref{eq:SpatialPerceptualError}:
\begin{align}
    \label{eq:ErrorImageSequence}
    \ErrorImageSequence(\SamplesSequence)
        &= \KernelTime \convolution \KernelSpace \convolution (\ImageSequence(\SamplesSequence) - \ReferenceSequence).
\end{align}
Since both reference and rendered images are subject to temporal perception, the convolution with this kernel is applied to both. Here, $\ErrorImageSequence(\SamplesSequence)$ denotes the sequence of per-frame error images.
In our experiments, we employ the kernel proposed by \citet{mantiuk2021fovvideovdp}. It is a sum of two components, a \emph{sustained} kernel and a \emph{transient} kernel, plotted in the inline figure. The sustained kernel encodes
{
\setlength{\columnsep}{4mm}%
\setlength{\intextsep}{4pt}%
\begin{wrapfigure}{r}{0pt}%
    \includegraphics{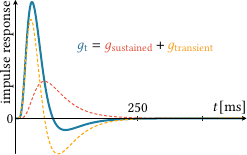}
\end{wrapfigure}%
 the response to slow temporal changes, and the transient kernel to fast changes~\cite{hammett1992two,burbeck1980spatiotemporal}. Note that in \cref{eq:ErrorImageSequence} the filter $\KernelTime$ is applied as a sliding window over the frames.

}

\paragraph{Temporal anti-aliasing}

TAA methods~\cite{yang2020surveytaa} compute pixel values as the weighted average of the current and previous frames. Such explicit filtering can be included in our model by expressing the image $\Image_i$ displayed at each frame as a convolution of the raw rendering results $\Estimate_j$ at all (past) frames: $\Image_i = [\KernelTAA \convolution \EstimateSequence]_i$. Substituting into \cref{eq:ErrorImageSequence}, the error-image sequence becomes
\begin{equation}
    \label{eq:SpatioTemporalTAAError}
    \ErrorImageSequence(\SamplesSequence)
        = \KernelTime \convolution \KernelSpace \convolution (\KernelTAA \convolution \EstimateSequence(\SamplesSequence) - \ReferenceSequence).
\end{equation}
In our experiments, we use an exponential moving average (EMA) kernel $\KernelTAA$, with weights $\KernelTAA(j) = \alpha (1 - \alpha)^j$, for $j \geq 0$, where $\alpha \in [0, 1)$ is a smoothing parameter (we use $\alpha = 0.2$).
Note here that the perceptual kernels $\KernelSpace$ and $\KernelTime$ are applied to both the image estimates and the reference image, but the TAA kernel $\KernelTAA$ is applied only to the raw image estimates.

\begin{figure}
    \centering
    \hspace*{-4.3mm}
    \includegraphics{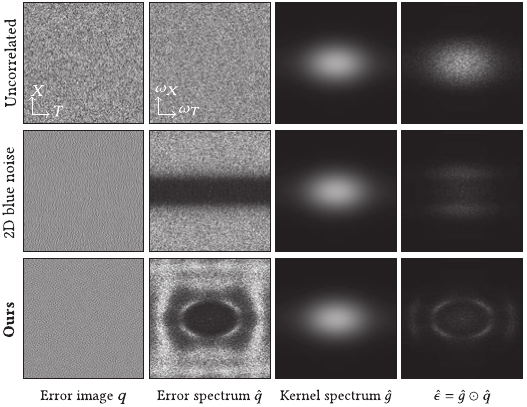}
    \caption{
        Spatio-temporal (\spatioTemporalDomain) slices of the error-image sequence (leftmost column) for white-noise (Uncorrelated), spatial-only blue-noise~\cite{salaun2022scalable} (2D blue noise), and spatio-temporal blue-noise (Ours) sample sets. The center two columns show the Fourier spectra of the error images (center left) and our perceptual kernel (center right). The rightmost column shows the product of these two (a.k.a.\ the perceptual error), \ie the Fourier spectrum of the error image convolved with the kernel. Our optimization minimizes the error spectrum $\hat{\ErrorImageSequence}$ (bottom row) and pushes the error outside of perceptible spatio-temporal frequency range (the window of visibility~\cite{Watson1986windowOfVisibility}; more details in \cref{par:discussion_spatiotemporal_model}).
    }
    \label{fig:error_kernel_product}
\end{figure}

\paragraph{Optimization objective}

Our objective is then to find the sample sequence $\SamplesSequence$ that minimizes the norm of the error-image sequence~\eqref{eq:SpatioTemporalTAAError}:
\begin{tcolorbox}[ams equation,after=,]
    \label{eq:OptimizationProblem}
    \SamplesSequence'
        \,=\, \argmin_{\SamplesSequence} \| \ErrorImageSequence(\SamplesSequence) \|.
\end{tcolorbox}
In our optimization algorithm, presented in the following section, we use the $L_1$ norm, \ie we find the sample sequence that minimizes the sum of absolute values of all error-image pixels over all frames.

\paragraph{Discussion}
\label{par:discussion_spatiotemporal_model}

We illustrate the impact of spatio-temporal kernel filtering in \cref{fig:error_kernel_product}, which provides a visual representation of the error image for three different methods, \ie sample sequences $\SamplesSequence$ (rows). The first column shows temporal slices of the raw error, \ie $q=\ImageSequence(\SamplesSequence) - \ReferenceSequence$, and the second column shows the power spectra of the discrete Fourier transform (DFT) of those raw-error slices. The last column shows the DFT spectra of the convolution of the error images with our spatio-temporal perceptual kernel $\Kernel = \KernelTime \convolution \KernelSpace$ (plotted in the second last column). Assuming that the viewing conditions and frame rate correspond to the kernels $\KernelSpace$ and $\KernelTime$, our spatio-temporal kernel $\Kernel$ approximates the window of visibility~\cite{Watson1986windowOfVisibility}. This window is defined in the frequency domain and its size is determined by the cut-off spatio-temporal frequencies. Signal outside the window is considered \emph{invisible} (imperceptible). By optimizing the sample sequence $\SamplesSequence$ to solve~\cref{eq:OptimizationProblem}, we not only reduce the magnitude of the spectrum, but also push the energy outside the window of visibility as much as possible. This reduces the residual perceived error for all visible spatio-temporal frequencies.

In summary, in this section we have presented a perception-driven  model~\eqref{eq:SpatioTemporalTAAError} to assess the spatio-temporal quality of a sample sequence. We model human perception by a series of convolutions, and (optionally) include explicit temporal filtering (TAA). In our main results we use this objective for a priori sample optimization, as discussed in the next section. \Cref{fig:priori_posteriori} shows that a posteriori optimization can benefit from this formulation, too.

Similarly to prior work on spatial-only error optimization \cite{heitz2019low,heitz2019distributing,chizhov2020perceptual}, we assume integrated (radiance) function to be locally smooth (Lipschitz continuous) in space and time. This smoothness assumption is essential for achieving a desirable outcome in the optimization process. The spatio-temporal CSF~\cite{kelly79motion,DALY98,mantiuk2022} further supports our assumption since the HVS is mostly sensitive towards low- to mid-frequency signals. In practice, this implies that the sampling quality is less relevant in regions where the smoothness assumption is not met.

\section{A priori optimization}
\label{secAprioriOptmization}

Our optimization problem~\eqref{eq:OptimizationProblem} is similar in structure to that of \citet{chizhov2020perceptual} who consider single-image optimization. This problem can be tackled in a priori or a posteriori manner (see \cref{sec:RelatedErrorDistribution}). We focus on a priori optimization due to its higher practical value of computing a sample set once that can be used on any scene. To that end, we extend the method of \citet{salaun2022scalable} to our spatio-temporal setting.

A priori methods assume that the ground-truth image is constant~\cite{georgiev2016blue,heitz2019distributing,belcour2021bluenoise}. We extend this assumption to the temporal domain. Convolving $\ReferenceSequence$ with the TAA kernel $\KernelTAA$ thus becomes a no-op that allows us to simplify our objective function~\eqref{eq:SpatioTemporalTAAError}: we combine all kernels into a single spatio-temporal kernel $\Kernel$:
\begin{align}
    \label{eq:SpatioTemporalTAAErrorApriori}
    \ErrorImageSequence(\SamplesSequence)
        &= \KernelSpace \convolution \KernelTime \convolution \KernelTAA \convolution (\EstimateSequence(\SamplesSequence) - \ReferenceSequence)
        = \Kernel \convolution (\EstimateSequence(\SamplesSequence) - \ReferenceSequence)
        .
\end{align}
In the a priori setting, both the raw sequence $\EstimateSequence(\SamplesSequence)$ and reference sequence $\ReferenceSequence$ are unknown, preventing the exact minimization of the error~\eqref{eq:SpatioTemporalTAAErrorApriori}. Instead, we aim to minimize an upper bound of that error.

\paragraph{Perceptual-error bound}

Under our perceptual model, the value of the $j$\textsuperscript{th} pixel in the $i$\textsuperscript{th} frame of the perceived raw sequence $\Kernel \convolution \EstimateSequence(\SamplesSequence)$ is an average of the responses of all samples, weighted by the kernel $\Kernel$ centered at $(i,j)$. \citet[Appendix D]{salaun2022scalable} derived a bound for the absolute error of weighted integral estimates, based on filtered optimal transport. In our case their bound reads
\begin{equation}
    \label{eq:PixelErrorBound}
    |\ErrorImageSequence_{i,j}(\SamplesSequence)|
        \,\leq\, L \! \int_{\mathbb{R}}W\left(\SamplesSequence_{\Kernel_{i,j}>\Slice}, \Density_{\Kernel_{i,j}>\Slice}\right) \dif\Slice .
\end{equation}
The bound assumes a smooth rendering function, \ie the incident radiance on the continuous image plane, with Lipschitz constant~$L$. It is an integral over Wasserstein distances $W$ between the optimized sample distribution $\SamplesSequence$ and the target (uniform) distribution $\Density$; these distributions are filtered to only include the mass at locations where the kernel value exceeds the threshold $\Slice$~\cite{salaun2022scalable}:
\begin{equation}
    \SamplesSequence_{\Kernel_{i,j}>\Slice} = \{ \Sample \in \SamplesSequence \mid \Kernel_{i,j}(\Sample) > \Slice \}
    .
\end{equation}
We use the 2-Wasserstein distance which is defined as
\begin{equation}
    W(\SamplesSequence, \Density) = \left( \inf_{\gamma \in \Gamma(\SamplesSequence,\Density)} \int_{\Omega^2} \|x - y\|^2 \,\dif \gamma(x,y) \right)^{\nicefrac{1}{2}}
    \text.
\end{equation}
Here $\Gamma(\SamplesSequence, \Density)$ is the set of all possible transport plans between the two distributions~\cite{bonnotte2013unidimensional}. Since the regular Wasserstein distance is difficult to compute, we further bound it via its sliced variant which involves only easy-to-compute 1D Wasserstein distances~\cite{Pitie:SlicedOptimalTransport}:
\begin{equation}
    W(\SamplesSequence,\Density) \, \leq \,
    SW(\SamplesSequence,\Density) = \! \int_{\mathbb{S}^{d-1}} W \left( \SamplesSequence^\theta,\Density^\theta \right) \dif \theta
    ,
\end{equation}
where $\SamplesSequence^\theta$ and $\Density^\theta$ are the projection of the sample set and the uniform density along the 1D line $\theta$. We provide more details on the Wasserstein error bound in Section~1 of the  supplemental document.

To bound the 1-norm of our objective function \eqref{eq:SpatioTemporalTAAErrorApriori}, we sum the error bounds~\cref{eq:PixelErrorBound} of all pixels $j$ in all frames $i$:
\begin{tcolorbox}[ams equation,after=,]
    \label{eq:SequenceErrorBound}
    \!\! \| \ErrorImageSequence(\SamplesSequence) \| = \! \sum_{i,j} \! |\ErrorImageSequence_{i,j}(\SamplesSequence)|
        \leq\, L \! \sum_{i,j} \! \int_{\mathbb{R}} \!\int_{\mathbb{S}^{d-1}} \!\!\!\!\! W\!\left(\SamplesSequence^\theta_{\Kernel_{i,j}>\Slice}, \, \Density^\theta_{\Kernel_{i,j}>\Slice}\right) \!\dif \theta \!\dif \Slice.\!\!
\end{tcolorbox}

\begin{figure}[t]
    \centering
    \includegraphics{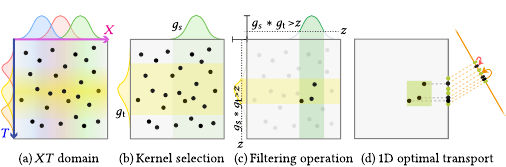}
    \vspace{-4.5mm}
    \caption{
        Visualization of how the gradients are estimated for our optimization. 
        Given the spatio-temporal ({\textcolor{magenta}{\spatialDomain}}{\textcolor{blue}{\temporalDomain}}) space and the kernels~(a,b), we randomly threshold their convolution to select a subset of samples~(c). The filtered sample set is then projected to a random 1D slice to compute the 1D-Wasserstein gradient~(d). The process is repeated multiple times to obtain a sufficiently low-noise gradient estimate.
    }
    \label{fig:StochasticMinimization}
\end{figure}

\paragraph{Gradient-descent optimization}

We minimize \cref{eq:SequenceErrorBound} via stochastic gradient descent, using Monte Carlo integration to estimate the involved integrals. We found that the Adam optimizer works best for our case, due to the sparse support of the kernels and the rather high noise of the gradient estimates. Details on the gradient computation can be found in supplemental Section 2. 

The process is illustrated in \cref{fig:StochasticMinimization}. At each optimization step, we first randomly select a kernel $\Kernel_{i,j}$ (\cref{fig:StochasticMinimization}b). We then sample a filtering threshold $\Slice$ which yields a sample subset $\SamplesSequence_{\Kernel_{i,j} > z}$ (\cref{fig:StochasticMinimization}c). Finally, a random slice $\theta$ is sampled to estimate the gradient of the sliced Wasserstein distance (\cref{fig:StochasticMinimization}d). This process is repeated multiple times to reduce the variance. The resulting multi-sample gradient estimate is then used to perform one gradient-descent update step. Algorithm~1 
summarizes these optimization steps.

\section{Results}
\label{sec:results}

We evaluate the rendering performance of our method by computing ray-traced direct illumination with PBRTv3~\cite{pharr2016physically}. We compare the results to the previous approach of \citet{wolfe2022stbn}, independent per-frame spatial-only blue noise~\cite{salaun2022scalable}, and the baseline of independent, white-noise sampling. Animations were rendered at 60Hz. Rendering is done using 1 sample per pixel unless stated otherwise.

We compute fixed-resolution spatio-temporal sample tiles, by toroidally wrapping the kernels during the optimization to ensure that they can be seamlessly tiled in space and time during rendering. If a spatial or temporal kernel has theoretically infinite support, we truncate it at the point where its values become negligible. We observe that a tile needs to be at least an order of magnitude larger than the truncated kernel to avoid tiling artifacts (supplemental, Fig.~1
). In our experiments, the kernel size is 7$\times$7 pixels and 8 frames wide (\ie 7$\times$7$\times$8 pixels). We found that a tile size of 128$\times$128$\times$30 pixels achieves the best trade-off between optimization cost and tiling artifacts. We use the same tile size for all methods.

\begin{pseudocode}[t]
    \caption{
        Our spatio-temporal sample optimization.
    }
    \label{alg:Temporal_FSOT_algo}
    \vspace{-1.5mm}
    \hrule height 0.7pt
    \vspace{0.5mm}
    \begin{algorithmic}[1]
    \Function{OptimizeSamples}{$IterationCount$, $BatchSize$}
            \State $\SamplesSequence = $ InitRandom() 
            \AlgCommentLeft{Initialize sample set}
            \State $Optimizer = $ InitAdamOpimizer()
            \For{$t = 1 .. IterationCount$}
                \State $g$ = $\textbf{0}$
                \For{$m = 1 .. BatchSize$} 
                    \State $k$ = SelectRandomKernel($M$)
                    \AlgCommentLeft{\cref{fig:StochasticMinimization}b}
                    \State $s$ = FilterSampleSet($\SamplesSequence$, $k$)
                    \AlgCommentLeft{\cref{fig:StochasticMinimization}c}
                    \State $g$ += EvaluateSWGradient(s)
                    \AlgCommentLeft{Accumulate gradient}
                \EndFor
                \State Update($Optimizer$, $g / BatchSize$)
            \EndFor
            \State \Return $\SamplesSequence$
        \EndFunction{}
    \end{algorithmic}
    \vspace{0.5mm}
    \hrule height 0.7pt
\end{pseudocode}

We computed the blue-noise tiles of \citet{wolfe2022stbn} using their public code. We slightly increased their spatial Gaussian kernel to a standard deviation of 2.1 (from 1.9), to match the spatial kernel used for all other methods. We used the public code of \citet{salaun2022scalable} to generate 30 independently optimized 2D blue-noise sample sets.

To mimic temporal perception in a static image, the renderings presented in the following are temporally pre-filtered with the kernel of \citet{mantiuk2021fovvideovdp} (see \cref{sec:SpatioTemporalError}). The visual quality of the results is best appreciated by referring to the supplemental video and HTML viewer.

For quantitative comparison, we compute the \pError, $\textstyle \nicefrac{\ErrorImageSequence_i^2(\Samples_i)}{(\Reference_i^2 \, + \, 0.01)}$, at the $i$\textsuperscript{th} frame. That is, we filter the rendered image and the reference according to our model~\eqref{eq:SpatioTemporalTAAError} and compute the relative MSE of the result for the desired frame ($16$\textsuperscript{th} frame, unless stated otherwise).

We apply our method to animation rendering with and without temporal anti-aliasing (TAA). For our method we optimize on sample set for each variant, tailored to the filter. \Cref{tab:Rendering_PRelMSE_Rafal} summarizes our quantitative results across a diverse set of test scenes. It shows the \pError of our method and previous works \cite{salaun2022scalable,wolfe2022stbn} relative to uncorrelated (\ie white-noise) spatio-temporal sampling. Across all scenes, our method consistently achieves better results, both with and without TAA.

\begin{table}
    \centering
    \caption{
        Perceptual error (\pError) across different scenes. The numbers are the ratio of the \pError of the different methods compared to the baseline of uncorrelated sampling; lower is better. Raw error values can be found in the supplemental document. We compare the methods with and without TAA. In both cases, and on every tested scene, our method achieves the lowest perceptual error. We set the standard deviation of Gaussian kernels to $\sigma=2.1$; for \citet{wolfe2022stbn} we also report results with $\sigma=1.9$ as used by them. 
    }
    \label{tab:Rendering_PRelMSE_Rafal}
    \setlength{\tabcolsep}{3pt}
    \small
    \scalebox{0.9}{
        \begin{tabularx}{1.10\columnwidth}{lcccccc}
            \toprule
            \textbf{Scene} &  \multicolumn{2}{c}{\footnotesize \textbf{\citet{salaun2022scalable}}} & \multicolumn{2}{c}{\footnotesize \textbf{\citet{wolfe2022stbn}}} & \multicolumn{2}{c}{\textbf{Ours}}\\
            ~ &   TAA & no TAA & TAA & no TAA & TAA & no TAA\\
            \midrule
            Chopper     & 0.61$\times$ & 0.62$\times$ & 0.69$\times$ \footnotesize (0.66$\times$) & 0.72$\times$ \footnotesize (0.69$\times$)  & \textbf{0.48$\times$} & \textbf{0.55$\times$} \\
            Teapot     & 0.65$\times$ & 0.63$\times$ & 0.80$\times$ \footnotesize (0.63$\times$) & 0.78$\times$ \footnotesize (0.65$\times$) & \textbf{0.56$\times$} & \textbf{0.58$\times$}\\
            Modern Hall~    & 0.90$\times$ & 0.85$\times$ & 0.98$\times$ \footnotesize (0.95$\times$) & 0.94$\times$ \footnotesize (0.91$\times$) & \textbf{0.87$\times$} & \textbf{0.83$\times$}\\
            Living room~  & 0.87$\times$ & 0.82$\times$ & 0.89$\times$ \footnotesize (0.86$\times$) & 0.86$\times$ \footnotesize (0.82$\times$) & \textbf{0.84$\times$} & \textbf{0.80$\times$}\\
            Dragon     & 0.54$\times$ & 0.52$\times$ & 0.66$\times$ \footnotesize (0.62$\times$) & 0.67$\times$ \footnotesize (0.63$\times$) & \textbf{0.48$\times$} & \textbf{0.51$\times$}\\
            Veach MIS & 0.87$\times$ & 0.83$\times$ & 0.99$\times$ \footnotesize (0.97$\times$) & 0.92$\times$ \footnotesize (0.90$\times$) & \textbf{0.72$\times$} & \textbf{0.69$\times$}\\
            \bottomrule
        \end{tabularx}
    }
    \vspace{2mm}
\end{table}

\paragraph{Direct viewing}

\Cref{fig:teaser,fig:comparison_wolfe_same_kernel} show results for animations without TAA using 1 sample per pixel. To aid interpretation of the results, the figures display the discrete Fourier transform (DFT) of different zoom-ins. Especially on the temporal slice in the bottom right of \cref{fig:teaser}, these show clearly where the improvements of our method stem from: While the previous approach of \citet{wolfe2022stbn} explicitly optimizes for 2D blue noise in image space and 1D blue noise along the temporal domain, our method optimizes samples for an exact kernel, dictated by a perception model. Consequently, the frequency distribution of the error with our method better matches the filter, resulting in a lower perceived error.

Our algorithm can be used to optimize sample sets for any number of samples per pixel. \Cref{fig:comparison_wolfe_4spp} shows an example using four samples. Compared to previous work, our approach achieves a better blue-noise distribution also at higher sample counts.

\paragraph{Temporal anti-aliasing}

The behavior with explicit TAA filtering is similar to that under direct viewing. Again, our method is consistently better than uncorrelated sampling, previous work~\cite{wolfe2022stbn}, and independent 2D blue noise~\cite{salaun2022scalable} across all test scenes (\cref{tab:Rendering_PRelMSE_Rafal}). \Cref{fig:Rendering_comparison_Perception_TAA} shows the TAA (and perception) filtered frames of two scenes. For our method, we compare two different optimization objectives: our full model and a simplified version where we left out the perception filter and only optimized for TAA. Optimizing only for TAA still outperforms previous work, but yields 5-10\% higher perceived error than utilizing the full model. This supports our hypothesis that optimizing for a more accurate kernel yields best results.

\section{Discussion}

\paragraph{Impact of kernel shape}

Our method differs from that of \citet{wolfe2022stbn} in two main aspects: the model and the optimization process. The approach of \citet{wolfe2022stbn} does not directly translate to a kernel in our optimization framework, since they separate the spatial and temporal dimensions. Therefore, to better understand how much of our improvements are due to the model, and how much due to the optimization itself, we performed an ablation where we optimized sample sets for kernels different from the one used for final filtering.

\Cref{tab:Rendering_PRelMSE_Rafal_TAA} summarizes the results. We report the error values for three different kernels: a symmetric Gaussian (standard deviation 2.1), the TAA kernel, and the full TAA and temporal perception kernel \cite{mantiuk2021fovvideovdp}. As expected, the best result is achieved when optimizing for the full model. Since the TAA and Gaussian kernels have similar shape, their results only differ by a few percent. These results indicate that, while matching the overall shape of the final kernel is important, exact match is not critical.

Note that all models in \cref{tab:Rendering_PRelMSE_Rafal_TAA} yield lower error than the approach of \citet{wolfe2022stbn}. This indicates that their separation into spatial and temporal components, while helpful for convergence in their optimizer, hampers the attainable quality.

\begin{table}
    \centering
    \caption{
        Ablation test for different optimization objectives. On an animation with TAA we test sample sets optimized for three temporal kernels: a symmetric Gaussian (standard deviation 2.1), an EMA TAA kernel, and our full model incorporating the TAA kernel and the perception kernel of \citet{mantiuk2021fovvideovdp}. The numbers are the relative reduction in perceived noise (\pError) compared to the method of \citet{wolfe2022stbn}; lower is better. Raw error values can be found in the supplemental document. Lowest error is achieved when the optimization is tailored to the full filter.
    }
    \setlength{\tabcolsep}{6pt}
    \begin{tabularx}{\columnwidth}{Xccc}
        \toprule
        \textbf{Scene} & \textbf{Gaussian} & \textbf{TAA} &  \textbf{Perception\,+\,TAA}\\
        \midrule
        Chopper & 0.75$\times$ & 0.75$\times$ &  \textbf{0.70$\times$}\\
        Teapot & 0.78$\times$ &  0.76$\times$ &  \textbf{0.70$\times$}\\
        Modern Hall & 0.92$\times$ &  0.90$\times$ &  \textbf{0.89$\times$}\\
        Living room & 0.97$\times$ &  0.96$\times$ &  \textbf{0.95$\times$}\\
        Dragon & 0.79$\times$ &  0.77$\times$ & \textbf{0.72$\times$}\\
        Veach MIS & 0.77$\times$ &  0.78$\times$ & \textbf{0.74$\times$}\\
        \bottomrule
    \end{tabularx}
    \label{tab:Rendering_PRelMSE_Rafal_TAA}
\end{table}

\begin{figure}[t!]
    \centering
    \hspace*{-0.8mm}
    \includegraphics[width=1.01\linewidth]{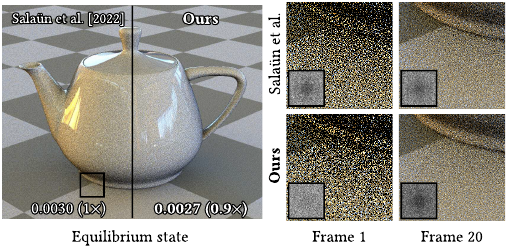}
    \vspace{-6mm}   
    \caption{
        Comparison between independent spatial-only optimization~\cite{salaun2022scalable} and our method on an animation with TAA. The image on the left is the 20\textsuperscript{th} frame, the zoom-ins show the state at the 1\textsuperscript{st} and 20\textsuperscript{th} frame, along with the DFT spectra of their error images. Spatial-only optimization is much better in the first frame, where no temporal filtering occurs. Our method shows better blue-noise quality once the steady state is reached.
    }
    \label{fig:independent_2D_vs_3D}
\end{figure}

\paragraph{Performance on the first frames}

Our optimization assumes that a sufficient number of past frames are available to apply the full temporal kernel. This is not the case early in an animation, as shown in \cref{fig:independent_2D_vs_3D}. The figure compares our result with spatial-only blue noise~\cite{salaun2022scalable} under environment-map illumination with TAA filtering. In the first frame, spatial-only optimization yields higher quality, since no temporal filtering can yet occur. In subsequent frames, our method performs better. This is also visible in the DFT spectra (insets) where our method has fewer low-frequency error components, \ie a better blue-noise distribution.

We extend this analysis by comparing the evolution of perceptual error with the number of frames. \Cref{fig:convergence_plots} shows this evolution on the Chopper, Dragon and Teapot scenes. The methods compared are uncorrelated sampling, \citet{salaun2022scalable}, \citet{wolfe2022stbn} and ours. The results show that at the start of each curve, when few frames are accumulated, spatial-only optimization performs best. This confirms the results presented above. However, as the frames accumulate and the equilibrium state is reached, our method obtains the lowest perceptual error.

\begin{figure*}[t!]
    \centering
    \hspace*{-1.5mm}
    \includegraphics{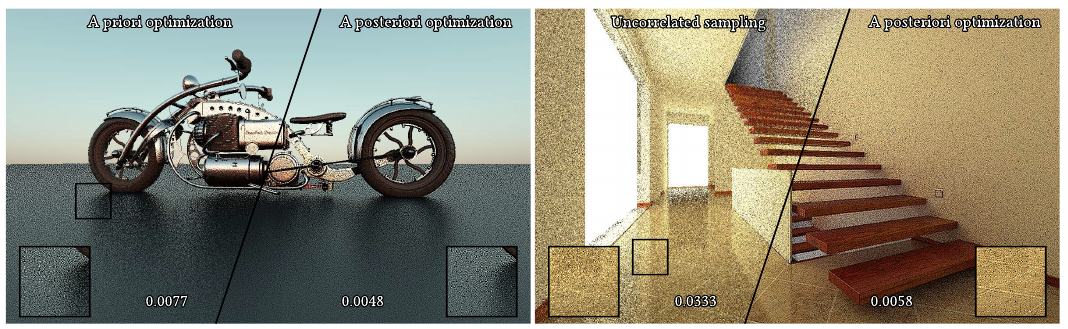}
    \vspace{-6mm}
    \caption{
        Our theory can be used to extend a posteriori perceptual error optimization \cite{chizhov2020perceptual}. Here we show improvement in direct-illumination (2D sampling) on the left and path tracing rendering (10D sampling) on the right. Unlike a priori methods, a posteriori optimization is not sensitive to sampling dimensionality and achieves higher quality thanks to image-based optimization. All images show the 16\textsuperscript{th} animation frame filtered with the temporal perception kernel of \citet{mantiuk2021fovvideovdp}, along with the corresponding \pError values.
    }
    \vspace{-1mm}
    \label{fig:priori_posteriori}
\end{figure*}

\paragraph{A posteriori optimization}

A priori optimization is inherently limited in the achievable quality because the optimized sample set must generalize to arbitrary scenes and importance-sampling transformations. A posteriori optimization can achieve better results, but a truly practical method has yet to be found. To explore the quality achievable by an a posteriori approach using our objective, we extended the method of \citet{chizhov2020perceptual} to the temporal domain, using our model. Specifically, we employ their ``vertical'' optimization which selects one out of 15 candidate samples for each pixel, to solve \cref{eq:OptimizationProblem}. We compare the result to our a priori optimization for the same kernel on the Chopper scene in \cref{fig:priori_posteriori}. Here, a posteriori optimization yields an notable improvement of 60\%. These results indicate that further research on (practical) a posteriori methods is worthwhile and can benefit from our formulation.

\paragraph{Optimization cost}

Our sample sets need only be computed once per filter kernel, number of integration dimensions, sample count, and frame rate. Nevertheless, when multiple variations of these parameters are desired, computation cost may become a concern. Our CPU implementation takes about 1--2 days to optimize one sample set using 10k SGD steps with a mini-batch size of 4k. The theoretical bottleneck is the computation of the 1D optimal transport, which relies on an $n \log(n)$ sorting operation per gradient-descent iteration. In practice, computation speed is significantly affected by accessing sample subsets that are scattered in memory. Furthermore, the use of a small subset of samples with non-zero gradients per step necessitates the use of large batch sizes, resulting in higher computation costs. Despite these challenges, parallelism can be harnessed within the algorithm: across different projections of the mini-batch on the CPU, also for sorting on the GPU. We believe that, with further performance improvements, \eg using a pre-optimized set as initialization, computation time can be reduced to at most a few hours per sample set.

\paragraph{Limitations}

The signal-constancy assumptions made by a priori perceptual error optimization hold only locally and approximately. In regions of large signal variation, \eg thin shadow penumbrae (spatially) or fast-moving objects (temporally), perceptual error increases. Temporal variations could be addressed via the use of motion vectors, which we leave for future work.

A significant limitation of a priori optimization methods lies in their applicability to more complex rendering algorithms, such as path tracing. This is due to the increased sampling dimensionality and the variation of the rendering function with longer paths. A priori optimization is not sensitive to dimensionality and can tailor the sampling to the rendered image.

The capacity of the error distribution to influence perceptual quality is also contingent upon the noise level present in the scene. When the noise level is low, the impact of the error distribution on perceptual quality diminishes.

\paragraph{Future work}

In this work we use a basic spatio-temporal CSF model \cite{chizhov2020perceptual,mantiuk2021fovvideovdp}, but our framework (\cref{secSpatioTemporalModel}) supports arbitrary filters. Exploring more advanced CSF models \cite{mantiuk2022} that account for display luminance (\eg darker displays increase the HVS tolerance for contrast errors and flickering, while saving energy) and foveation (sparser sampling with increasing retinal eccentricity) could be a way to further improve visual quality. A hold-type blur of moving objects that arises in the HVS as a function of display persistence and refresh rate~\cite{jindal2021perceptual} can also lead to increasing the HVS tolerance to rendering error. One could specifically optimize sampling by considering content-dependent visual masking \cite{mantiuk2021fovvideovdp}, \eg by precomputing a texture-specific sample set.

Another promising future direction would be to find a practical approach for a posteriori optimization of sample patterns. Ideally, such a method would work in real-time applications and complex light-transport algorithms.

The relationship between sample optimization and denoising is another interesting topic. As noted by \citet{heitz2019distributing} and \citet{chizhov2020perceptual}, high-frequency error distributions can afford higher fidelity when denoising via low-pass filtering; existing denoisers may need adjustment or retraining to optimally handle such input. We believe that our high-frequency spatio-temporal distribution paves the way for devising improved, correlation-aware interactive denoising methods.

\section{Conclusion}

We have introduced a general model and a practical method for spatio-temporal sample optimization for Monte Carlo animation rendering. Our method accounts for both perceptual and explicit temporal filtering. To achieve practicality, we extend an existing a priori optimization method to support our spatio-temporal model. As a result, we can precompute scene-agnostic sample sets that yield considerable improvements over previous work in terms of perceived noise quality.

\section*{Acknowledgements}

This project has received funding from the European Union’s Horizon 2020 research and innovation program under \href{https://prime-itn.eu}{Marie Sk\l{}odowska-Curie} grant agreement \textnumero 956585. We thank the anonymous reviewers for their feedback, and the authors of the following scenes: julioras3d (Chopper), NewSee2l035 (Modern Hall), Benedikt Bitterli (Utah Teapot), Wig42 
(Living Room), and Jay Hardy (White Room).


\bibliographystyle{ACM-Reference-Format}
\bibliography{paper}


\begin{figure*}[t!]
    \centering
    \hspace*{-2.6mm}
    \includegraphics[width=1.016\textwidth]{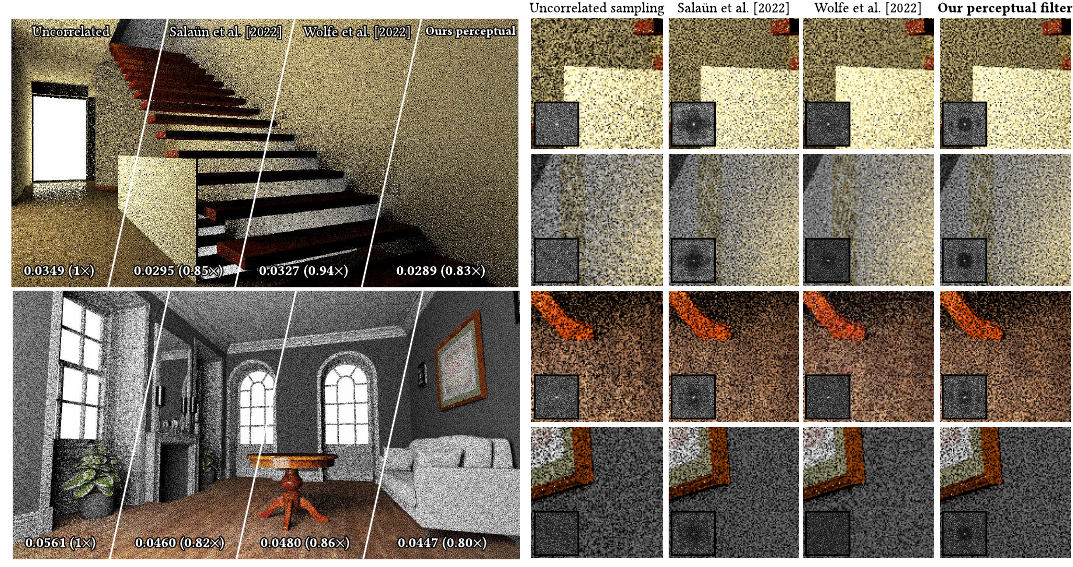}
    \vspace{-7mm}
    \caption{
        Comparison of the 16\textsuperscript{th} animation frame, without TAA. To mimic human perception, for display we apply the temporal filter of \citet{mantiuk2021fovvideovdp}. We compare our method to uncorrelated sampling and the methods of \citet{salaun2022scalable} and \citet{wolfe2022stbn}. The insets in each crop show the DFT of the error image, and the numbers on the left are the \pError of each method (lower is better). We achieve visible improvements over previous work on all scenes, and a more pronounced blue-noise distribution in the DFT spectrum.
        Please refer to the supplemental HTML viewer to better appreciate the differences.
    }
    \label{fig:comparison_wolfe_same_kernel}
\end{figure*}

\begin{figure*}[t!]
    \centering
    \hspace*{-1.5mm}
    \includegraphics[width=1.007\textwidth]{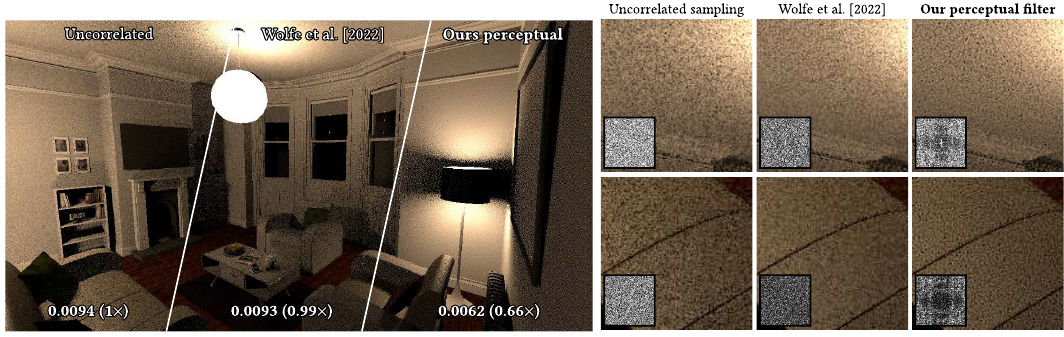}
    \vspace{-6mm}
    \caption{
        Rendering comparison on the 10\textsuperscript{th} animation frame, rendered with 4 samples per pixel. To mimic human perception, for display we apply the temporal filter of \citet{mantiuk2021fovvideovdp}. We compare our method to uncorrelated sampling and the method of \citet{wolfe2022stbn}. The insets in each crop show the DFT of the error image, and the numbers on the left are the \pError of each method (lower is better). Our method preserves the desirable blue-noise error distribution also at higher sample counts.
    }
    \label{fig:comparison_wolfe_4spp}
\end{figure*}

\begin{figure*}[t!]
    \centering
    \includegraphics{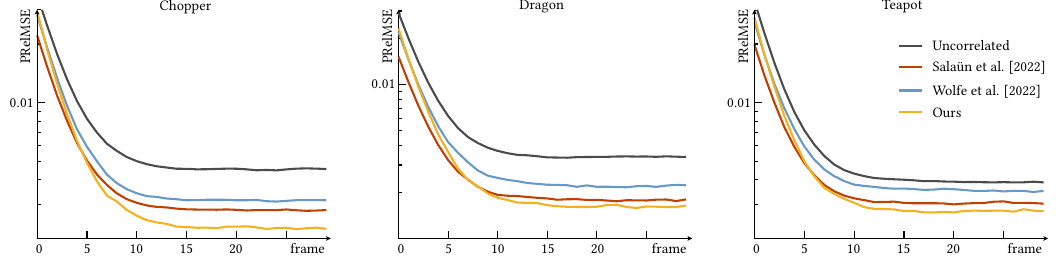}
    \vspace{-3mm}
    \caption{
        Perceptual error across the first 30 animation frames (without motion) for three scenes rendered with TAA. The error initially reduces for all methods, as more frames are included in the temporal filter; until frame 15, where the full support of the kernel is reached. Spatial-only blue noise~\cite{salaun2022scalable} performs best for the first few frames, where not much temporal filtering yet occurs. Our optimization achieves the lowest perceptual error at the steady state.
    }
    \label{fig:convergence_plots}
\end{figure*}

\begin{figure*}[t!]
    \centering
    \vspace{5mm}
    \hspace*{-1.2mm}
    \includegraphics[width=1.008\textwidth]{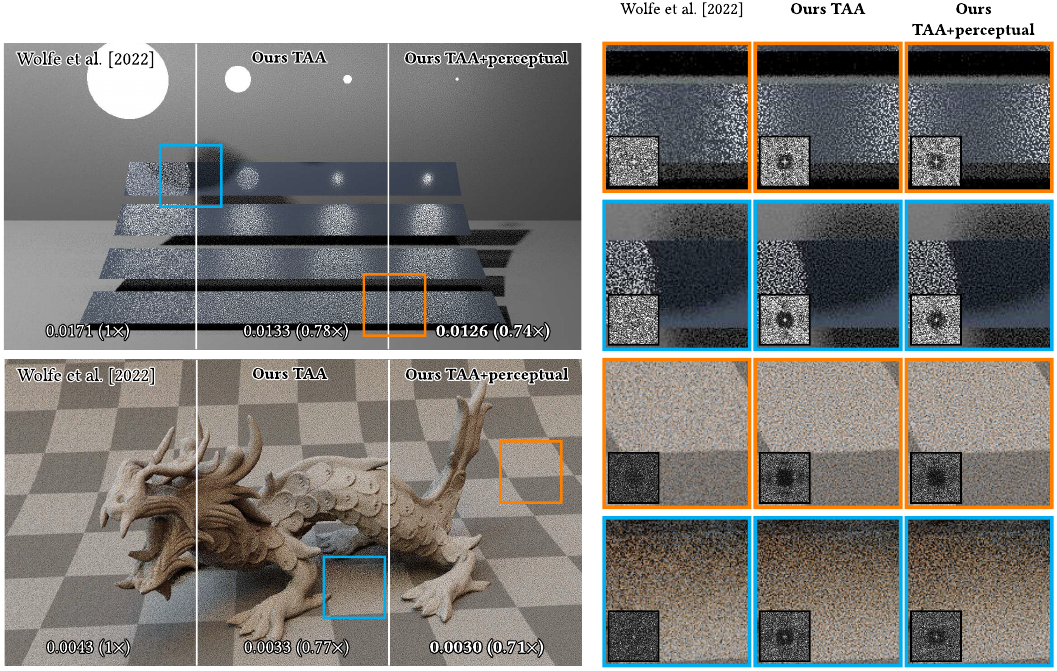}
    \vspace{-6mm}
    \caption{
        Rendered images with temporal anti-aliasing (TAA). We mimic human perception by applying the temporal filter of \citet{mantiuk2021fovvideovdp}. As an ablation, we compare our method optimized only for the TAA kernel (center) and the full result (right) to previous work. We provide 2 crops for each scene associated with the DFT of the error in the region. The numbers at the bottom are the \pError for 16\textsuperscript{th} frame for each method, lower is better. Optimizing only for TAA already performs well, but optimizing for both TAA and perception yields best results.
    }
    \label{fig:Rendering_comparison_Perception_TAA}
\end{figure*}


\end{document}